\documentclass[12pt, amsfonts, amssymb]{article}

\usepackage{amssymb}
\usepackage{amsmath}

\newcommand{\CC}{{\mathbb C}}
\newcommand{\RR}{{\mathbb R}}
\newcommand{\ZZ}{{\mathbb Z}}
\newcommand{\PP}{{\mathbb P}}

\newcommand{\NN}{{\mathbb N}}
\renewcommand{\SS}{{\mathbb S}}

\newcommand{\Tr}{{\rm Tr}}
\newcommand{\diag}{{\rm diag}}
\newcommand{\ra}{\rightarrow}
\newcommand{\tA}{{\tilde{A}}}
\newcommand{\V}{{\vec{V}}}
\newcommand{\vA}{{\vec{A}}}
\newcommand{\vtA}{{\vec{\tilde{A}}}}

\newcommand{\nn}{\nonumber}

\newcommand{\hphi}{{\hat\phi}}
\newcommand{\hA}{{\hat A}}
\newcommand{\bpartial}{{\bar\partial}}
\newcommand{\hF}{{\hat F}}
\newcommand{\bs}{{\bar s}}
\newcommand{\bz}{{\bar z}}
\newcommand{\vr}{v_{{\rm ren}}}

\title{\bf Hyperk\"ahler Metrics from\\ Periodic Monopoles}
\author{
Sergey A. Cherkis\thanks{e-mail: cherkis@physics.ucla.edu}\\
\it UCLA Physics Department, Los Angeles, CA 90095-1547, USA
 \rm
\and
Anton Kapustin\thanks{e-mail: kapustin@theory.caltech.edu}\\
\it California Institute of Technology, Pasadena, CA 91125, USA
}

\begin{document}
\begin{titlepage}

\renewcommand{\thepage}{ }
\date{}

\maketitle

\begin{abstract}
Relative moduli spaces of periodic monopoles provide novel examples of Asymptotically Locally Flat hyperk\"ahler manifolds. By considering the interactions between well-separated periodic monopoles, we infer the
asymptotic behavior of their metrics. When the monopole moduli space is
four-dimensional, this construction yields interesting examples of metrics
with self-dual curvature (gravitational instantons).
We discuss their topology and complex geometry.
An alternative construction of these gravitational instantons using moduli spaces of Hitchin equations is also described.

\end{abstract}
\vspace{-5.5in}

\parbox{\linewidth}
{\small\hfill \shortstack{CALT-68-2347\\ UCLA/01/TEP/20 \\ CITUSC/01-031}}

\end{titlepage}
\pagestyle{headings}

\section{Introduction}

One of the most powerful methods for obtaining hyperk\"ahler manifolds is the
Hyperk\"ahler Quotient Construction~\cite{HKLR}. Most known hyperk\"ahler
manifolds are hyperk\"ahler quotients of affine hyperk\"ahler spaces by a suitable subgroup of tri-holomorphic isometries. For example, all Asymptotically Locally Euclidean four-dimensional hyperk\"ahler manifolds (in other words, ALE gravitational instantons) have been constructed in this way~\cite{Kr}. The affine hyperk\"ahler space is finite-dimensional in
this case.

More general hyperk\"ahler manifolds are obtained if one starts with an
infinite-dimensional affine hyperk\"ahler space and quotients by an
infinite-dimensional subgroup of isometries. Well-known examples
of this sort are moduli spaces of instantons on $\RR^4$
and moduli spaces of monopoles on $\RR^3$. The affine space is the space
of connections on a vector bundle on $\RR^4$ in the first instance, and the space of pairs $(connection,Higgs\ field)$ in the second instance. The quotienting group is the group of gauge transformations in both instances.

The monopole example is particularly nice, as one can determine the asymptotic
behavior of the metric from simple physical considerations~\cite{Mtn,GM}.
In the asymptotic region the monopoles are well separated, and can be regarded
as point particles interacting via long-range scalar and electromagnetic
fields. Each particle has an internal degree of freedom living on a circle,
which when excited gives the monopole an electric charge (i.e. makes it into
a dyon). In the asymptotic region the radius of this circle is a fixed
number determined by the vacuum expectation value of the Higgs field at
infinity. It follows that asymptotically the moduli space of $k$ $SU(2)$
monopoles looks like a $T^k$ fibration over $(\RR^3)^k/S_k$, where we divided
by the symmetric group $S_k$ to take into account the indistinguishability
of monopoles. Since the electric charges are conserved, the
fiberwise action of $T^k$ must be an isometry (in fact, a tri-holomorphic isometry).
A more detailed
analysis of the long-range interactions of moving monopoles yields
the precise metric in the asymptotic regime~\cite{Mtn,GM}, which turns out
to be Asymptotically Locally Flat.

It is customary to quotient the moduli
space by the translations of $\RR^3$ and the diagonal of $T^k,$ or equivalently to fix the center-of-mass coordinates of the monopoles
and the sum of their internal degrees of freedom (phases). The resulting $4(k-1)$-dimensional manifold is again hyperk\"ahler and is called the
relative (or centered) moduli space. The relative moduli
space of two monopoles is known as the Atiyah-Hitchin manifold~\cite{AH}. At infinity it looks
like a circle of fixed radius fibered
over $\RR^3/\ZZ_2,$ and the asymptotic metric has the Taub-NUT form.

One can generalize this example somewhat and consider $SU(2)$ monopoles moving in a background of $n$ point-like Dirac monopoles sitting at fixed locations~\cite{usthreed}. If the number of $SU(2)$ monopoles is one,
then the (uncentered) moduli space is the multi-Taub-NUT
space~\cite{KrTh,usthreed}.
It is a four-dimensional ALF manifold with a tri-holomorphic $U(1)$ isometry
isomorphic as a complex variety to a blow-up of $\CC^2/\ZZ_n.$
At infinity it looks like a circle of fixed radius fibered over $\RR^3$,
and the $U(1)$ action is fiberwise.\footnote{The multi-Taub-NUT metric can also be obtained by a finite-dimensional HKQ construction~\cite{HKLR}, or by using the Gibbons-Hawking ansatz~\cite{GH}.}
If the number of $SU(2)$ monopoles is two, then the
{\it relative} moduli space is four-dimensional and ALF, but does not have a
tri-holomorphic
$U(1)$ isometry. As a complex variety the moduli space is isomorphic to
a blow-up of $\CC^2/\Gamma$, where $\Gamma$ is a binary dihedral
group~\cite{usthreed}.
The asymptotic metric has the Taub-NUT form and looks at infinity like a circle of fixed radius fibered over $\RR^3/\ZZ_2$. In particular the
asymptotic metric has a tri-holomorphic $U(1)$ isometry which acts
fiberwise.

ALE gravitational instantons also have an asymptotic tri-holomorphic $U(1)$ isometry, but the circumference of the orbits grows
linearly as a function of the ``radius.'' Finite-dimensional HKQ construction
suffices to construct all such manifolds. In the cases when the
circumference of the orbits stays fixed at infinity, one needs to resort
to the infinite-dimensional HKQ construction, in general.

An obvious generalization is to consider ALF gravitational instantons
which asymptotically have a tri-holomorphic $T^2$ action. We will call such
gravitational instantons {\it ALG manifolds}. Such manifolds previously
arose in the physics literature as quantum moduli spaces of $d=4$ $N=2$ gauge
theories compactified on a circle (see below). No ``classical''
construction of such manifolds has been known previously. In this paper we
will produce examples of ALG manifolds using an infinite-dimensional HKQ
construction. More generally, we will show
how to construct ALF hyperk\"ahler manifolds of dimension $4(k-1)$ which
asymptotically have a tri-holomorphic $T^{2(k-1)}$ isometry.
To this end we will consider $k$ $SU(2)$ monopoles on
$\RR^2\times\SS^1$ with a flat metric. Such ``periodic'' monopoles have
been studied in Refs.~\cite{usone, ustwo}.
It was shown there that although each
periodic monopole has a logarithmically divergent mass, the relative
moduli space has a well-defined hyperk\"ahler metric. We expect
that this metric is smooth and geodesically complete. The asymptotic
behavior of this metric will be determined along the lines of Refs.~\cite{Mtn,GM}.
We will also consider a more general problem of periodic $SU(2)$ monopoles
moving in a background of point-like Dirac monopoles.

In the case $k=2$ the moduli space is four-dimensional, and
we will describe its geometry in some detail using the results of Refs.~\cite{usone,ustwo}. In fact, since the number of Dirac singularities
$n$ can vary from $0$ to $4,$ in this way we obtain five topologically
distinct four-dimensional hyperk\"ahler manifolds. We show that they are
ALG manifolds. Moreover, we will see that
the moduli spaces have a distinguished complex structure in which they
look like elliptic fibrations over $\CC$. The volume of the elliptic
fiber is constant in the asymptotic region of the moduli space.
The asymptotic $T^2$ isometry
acts on the fibers in a natural manner. The number and type of singular
fibers depends on the parameters of the metric.
We will discuss which kinds of singular fibers occur,
compute the Betti numbers of the moduli spaces, and in some cases the intersection pairing on the second homology.
We will see that the most general ALG gravitational instanton one can get in this way has an intersection form which is the affine Cartan matrix of type $D_4$. All other gravitational instantons we construct can be regarded as its degenerations.

Finally, we explain an alternative construction of our ALG manifolds
using moduli spaces of Hitchin equations~\cite{HitchinSpec} on a cylinder.
The two constructions are related by a version of Nahm transform~\cite{usone,ustwo}.

As discussed in Refs.~\cite{usone,ustwo}, moduli spaces of
periodic monopoles are closely related to certain $N=2,$ $d=4$
quantum gauge theories. For example, the moduli space of $k$
$SU(2)$ monopoles moving in a background of $n$ Dirac monopoles
is isomorphic to the quantum Coulomb branch of $SU(k)$ gauge
theory with $n$ fundamental hypermultiplets compactified on a
circle. The $D_4$ gravitational instanton mentioned above
corresponds to the $SU(2)$ gauge theory with four
hypermultiplets, while its degenerations correspond to the
$SU(2)$ gauge theory with three or fewer hypermultiplets. The
quantum Coulomb branch of these theories on $\RR^4$ has been
determined in two celebrated papers by Seiberg and
Witten~\cite{SW1,SW2}. Our results provide information about the
same theories on $\RR^3\times\SS^1$. The asymptotic form of the
metric on the Coulomb branch has been computed in~\cite{SW3, JWKV, WKV}.
The result agrees with the asymptotics of the metric
on the moduli space of periodic monopoles computed below.
However, if one want the complete metric, the gauge theory realization
is not very useful, since the metric is corrected by gauge theory
instantons. Such non-perturbative effects lead to exponentially small
corrections to the metric which are quite hard to compute.
On the other hand, we realized the same manifolds as classical objects,
namely as moduli spaces  of Bogomolny or Hitchin equations.
We hope that the corresponding hyperk\"ahler metrics can be
computed in a closed form using twistor methods.

\section{Asymptotic Metric on the Monopole Moduli Spaces}
\subsection{Generalities}
We will use the conventions of Refs.~\cite{usone,ustwo}.
We identify $\RR^2\times \SS^1$ with $\CC\times \SS^1$ and use a complex affine coordinate
$z$ on $\CC$
and a real coordinate $\chi$ on $\SS^1$ with an identification
$\chi\sim \chi +2\pi.$
For monopoles located at points $a_j=(z_j,\chi_j),\ j=1,\ldots,k,$
the field configuration at a distant point $x=(z,\chi)$ is given in a
suitable gauge by
\begin{eqnarray}\label{first}
\phi(x)&=&v+\sum_{j=1}^k \phi^j(z-z_j),\\
A_z&=&0,\ \ A_{\chi}=b+\sum_{j=1}^k A^j_{\chi}(z-z_j).
\end{eqnarray}
When all the distances $|z_i-z_j|$ are large, we interpret these fields as a superposition of the background fields, given by constants
$b$ and $v$, and individual fields of the monopoles $\phi^j$ and $A^j$.

When all monopoles are well separated, it is natural to think of their
dynamics in terms of motion and interaction of particles on $\RR^2\times\SS^1$. The moduli space coordinates are understood as parameterizing the positions of these $k$ particles as well as their internal degrees of freedom valued in $\SS^1$
(phases). A particle whose phase is changing with time
aquires an electric charge proportional to the rate of the phase
change~\cite{Mtn}. This is consistent with charge conservation because the
rate of phase change is an integral of motion.
Motion on the moduli space is thus interpreted as motion of $k$ dyons on $\RR^2\times\SS^1$.

So far the discussion parallels that for monopoles on $\RR^3$~\cite{Mtn}.
But unlike for monopoles on $\RR^3,$ there is a subtlety here related to the
fact that a single periodic monopole has infinite mass, because the
integral of the energy density logarithmically diverges at long distances
\cite{usone, ustwo}. One might conclude that the kinetic energy
associated with the motion on the moduli space is infinite as well.
If this were the case, the metric on the moduli space would be ill-defined
(divergent), and the positions of the particles would be parameters rather
than moduli. In fact, only the coordinates of the center of mass and the total
phase are parameters. The kinetic energy of the relative motion is finite,
and therefore there is a finite metric on the relative moduli
space~\cite{ustwo}.
To deal with this subtlety, we use the following procedure. In terms of the universal covering space of $\RR^2\times\SS^1,$ each periodic monopole is an array of infinitely many 't~Hooft-Polyakov monopoles. Such an array has an
infinite mass per unit length because of the divergence mentioned above. We regularize the problem by replacing
each infinite array by a finite array of $2N+1$ monopoles. This way all the masses and fields are finite. At the end of the computation we will send $N$ to infinity. As a result, we indeed recover a finite metric on the relative moduli space and verify that the center of mass and total phase of the configuration are parameters (the kinetic energy associated with them
diverges logarithmically as $N\ra\infty$).

With this remark in mind, the Higgs field produced by one periodic monopole of charge $g$ located at $z=0$ at distances large compared to the size of the monopole is
\begin{equation}
\phi^j(x)=\sum_{l=-N}^N\frac{-g}{\sqrt{|z|^2+(\chi-2\pi l)^2}}.
\end{equation}
We note for future use that for an 't~Hooft-Polyakov monopole $g=1,$ and for a singular Dirac monopole $g=-1/2$~\cite{ustwo}.
Since we are going to send $N$ to infinity, we may assume that $|z|\ll N.$
In this region the expression for $\phi^j$ simplifies:
\begin{equation}
\phi^j(x)=\frac{g}{\pi}\log|z|-gC_N+O\left(\frac{1}{|z|}\right).
\end{equation}
Here $C_N$ a positive constant diverging logarithmicaly with $N$; it will eventually be absorbed into the constant background $v$. From now on we shall omit terms decaying as $1/|z|$ or faster when writing the monopole fields.

The connection $A^j$ corresponding
to $\phi^j(x)$ is given by
\begin{equation}
A^j_{\chi}=\frac{g}{\pi}\arg z, \ \ A^j_z(x)=0,
\end{equation}
in a suitable gauge. To be precise, we should have added an $N$-dependent
constant to $A^j_\chi,$ but since it can be absorbed into the constant
background $b,$ we did not write it explicitly.

For convenience we define two auxiliary functions:
$$u(z)=\frac{1}{\pi}\log|z|-C_N, \ \ w(z)=\frac{1}{\pi}\arg z.$$

Note that the total field $\phi(x)$
of Eq.(\ref{first}) is given for large $z$ by
\begin{equation}
\phi(x)=v-kg C_N+\frac{kg}{\pi}\log|z|.
\end{equation}
Since we are interested in field configurations with fixed asymptotics,
it is natural to introduce
$$\vr=v-kg C_N$$
and adjust $v$ so that $\vr$ remains fixed when $N$ is sent to infinity.
Thus in the limit $N\ra +\infty$ we have $v\ra +\infty.$

Next we introduce an electric charge $q_j$ for each monopole. The resulting dyons acquire new interactions.
The Lagrangian of the $k$-th dyon is
\begin{equation}
L_k=-4\pi\phi\sqrt{g^2+q_k^2}\sqrt{1-\V_k^2}+4\pi q_k\V_k\cdot\vA-4\pi
q_k A_0+4\pi g\V_k\cdot\vtA-4 \pi g\tA_0.\nonumber
\end{equation}
Here $\V$ is the velocity 3-vector of the $k$-th dyon with components
$({\rm Re}\,\dot{z}, {\rm Im}\,\dot{z},\dot{\chi})$ (dot denotes time derivative). The fields
$\phi, A,$ and $\tA$ are superpositions of the fields
produced by other dyons evaluated at the location of the $k$-th dyon and
the constant background.
The magnetic charge $g$ couples to the ``magnetic'' potentials $\vtA, \tA_0$
which are dual to the ``electric'' potentials $\vA, A_0$ and are defined by
\begin{equation}
\begin{array}{rcccccl}
\nabla\times\vtA & \equiv &\vec{\tilde{B}} & = & -\vec{E} & \equiv &\nabla A_0+\frac{\partial}{\partial t}{\vA},\\
-\nabla\tA_0-\frac{\partial}{\partial t}\vtA &\equiv
&\vec{\tilde{E}}&=&\vec{B}&\equiv &\nabla\times\vA.
\end{array}
\end{equation}

The fields produced by a dyon at rest located at $z=0$ are
\begin{equation}\label{dyon}
\phi^j(x)=\sqrt{g^2+q_j^2}\ u(z),
\end{equation}
\begin{equation}
\begin{array}{lll}
A^j_{\chi}(x)=g w(z), &A^j_0(x)=-q_j u(z), &A^j_z(x)=0,\\
\tA^j_{\chi}(x)=-q_j w(z), &\tA^j_0(x)=-g u(z), &\tA^j_z(x)=0.
\end{array}
\end{equation}

The fields of a moving dyon are obtained by a Lorentz boost. Keeping terms
up to second order in velocities in $\phi,A_0,\tA_0$ and up to first
order in $\vA,\vtA,$ we get:
\begin{eqnarray}
\phi^j(x)&=&\sqrt{g^2+q_j^2}u(z)\sqrt{1-\V_j^2},\nn\\
A^j_{\chi}(x)&=&-q_j u(z)V_{j\chi}+g w(z),\nn\\
A^j_z(x)&=&-q_ju(z) V_{jz}\nn\\
A^j_0(x)&=&-q_j u(z)+g w(z) V_{j\chi},\\
\tA^j_{\chi}(x)&=&-g u(z)V_{\chi}-q_j w(z),\nn\\
\tA^j_z(x)&=&-g u(z)V_{jz},\nn\\
\tA^j_0(x)&=&-g u(z)-q_j w(z) V_{j\chi}. \nn
\end{eqnarray}
Following Ref.~\cite{GM}, we omitted certain terms of second order in
velocities in $\phi,\tA_0$
by replacing $1/\sqrt{r^2-\left({\bf r}\times {\bf V}\right)^2}$
with $1/r$ in the Li\'{e}nard-Wiechert potentials. This is allowed because
such second-order terms enter the kinetic energy with the same coefficients
as $1/r$ terms enter the static energy. Since the static interactions
cancel, so do the second-order terms of this type.

\subsection{Two-Monopole Interactions}\label{2mon}

Consider the Lagrangian of the $k$-th dyon in the presence of a dyon with
$j=1$.
Keeping terms up to second order in electric charges and velocities, we get
\begin{eqnarray}\label{one}
L_k&=&-m_k+\frac{1}{2}m_k \V_k^2+2\pi g^2 u(z_k-z_1) \left(\V_k-\V_1\right)^2+\nn\\
&&+4\pi g w(z_k-z_1)\left(q_k-q_1\right)\left(V_{k\chi}-V_{1\chi}\right)-\\
&&-2\pi u(z_k-z_1)\left(q_k-q_1\right)^2+4\pi b q_k V_{k\chi}.\nn
\end{eqnarray}
Here the dyon's rest mass $m_k$ is given by $4\pi v\sqrt{g^2+q_k^2}.$
Expanding $m_k$ to second order in $q_k,$ omitting a constant term
$-4\pi g v,$ and symmetrizing with respect to the two particles,
we obtain the total Lagrangian for the dyons with $j=1$ and $j=k:$
\begin{eqnarray}\label{two}
\frac{1}{4\pi}L_{1k}&=&-\frac{v}{2g}q_1^2-\frac{v}{2g}q_k^2+\frac{gv}{2} \V_1^2+\frac{gv}{2} \V_k^2+\frac{g^2}{2} u(z_k-z_1)(\V_k-\V_1)^2\nn\\
&&+\left(\frac{b}{2}+g w(z_k-z_1)\right)(q_k-q_1)(V_{k\chi}-V_{1\chi})-\\
&&-\frac{1}{2} u(z_k-z_1)(q_k-q_1)^2+
\frac{b}{2}(q_1+q_k)(V_{1\chi}+V_{k\chi}).\nn
\end{eqnarray}

Hence the Lagrangian describing the relative motion of the two dyons is
\begin{eqnarray}\label{interaction}
\frac{1}{4\pi}L_{\rm rel}&=&g^2(\frac{\vr}{4g}+\frac{1}{2\pi}
\log|z_k-z_1|)(\V_k-\V_1)^2+\\
&&+\left(\frac{b}{2}+\frac{g}{\pi}\arg(z_k-z_1)\right)(q_k-q_1)
(V_{k\chi}-V_{1\chi})-\nn\\
&&-\left(\frac{\vr}{4g}+\frac{1}{2\pi}\log|z_k-z_1|\right)(q_k-q_1)^2,\nn
\end{eqnarray}
while the Lagrangian for the motion of the center of mass is
$$
\frac{1}{4\pi}L_{CM}=\frac{vg}{4}\left(\V_1+\V_k\right)^2-
\frac{v}{4g}\left(q_1+q_k\right)^2+\frac{b}{2}\left(q_1+q_k\right)
\left(V_{1\chi}+V_{k\chi}\right).
$$
In the limit $N\ra\infty,\ v\ra\infty$ with $\vr$ and $b$ fixed, the relative Lagrangian stays finite, while the center-of-mass Lagrangian diverges, as expected.

Now we have to extract from $L_{\rm rel}$ the effective metric on the
relative moduli space. As explained above, the electric charges $q_j$
are conserved momenta conjugate to phase degrees of freedom $t_j$ associated
to monopoles. Since the monopoles are indistinguishable, we may assume that
$t_j$ are periodic variables with the same period. The coordinates on the relative moduli space of two monopoles are $z=z_k-z_1$, $\chi=\chi_k-\chi_1,$ and $t=t_k-t_1.$ To read off the metric on the
moduli space, we need to reintroduce the dependence on $\dot{t_j}$ into
the Lagrangian. This is achieved by the Legendre transform with respect
to $q=q_k-q_1.$ We let
$$
L'_{\rm rel}=L_{\rm rel}+4\pi g\,\dot{t} q,
$$
solve the algebraic equation of motion for $q,$ and substitute back into
$L'_{\rm rel}.$ The factor $4\pi g$ in front of $\dot{t}$ is introduced
for convenience. The result is
\begin{equation}\label{Lreltwo}
\frac{1}{4\pi g^2}L'_{\rm rel}= \frac{\tau_2(z)}{2}\left(|\dot{z}|^2+\dot{\chi}^2\right)
+\frac{1}{2\tau_2(z)} \left(\dot{t}+\tau_1(z)\dot{\chi}\right)^2,
\end{equation}
where
\begin{eqnarray}
\tau_1(z)&=&\frac{b}{2g}+\frac{1}{\pi}\arg(z),\\
\tau_2(z)&=&\frac{\vr}{2g}+\frac{1}{\pi}\log|z|.
\end{eqnarray}
{}From now on and to the end of this subsection we set $g=1,$ as appropriate
for non-abelian monopoles.

{}From Eq.~(\ref{Lreltwo}) we read off the asymptotic
metric on the moduli space. Setting
\begin{equation}\label{tautwomon}
\tau(z)=\tau_1(z)+i\tau_2(z)=\frac{i}{2}(\vr-ib) + \frac{i}{\pi}\log\bz,
\end{equation}
we can write the metric as follows:
\begin{equation}\label{elliptic}
\frac{1}{4\pi} ds^2=\tau_2(z) |dz|^2+\frac{1}{\tau_2(z)}\left|dt+\tau(z)d\chi
\right|^2.
\end{equation}

Eq.~(\ref{elliptic}) is a special form of the Gibbons-Hawking ansatz~\cite{GH}
which depends on a harmonic function on $\RR^3$. In our case the
harmonic function is $\tau_2(z)$. It is well-known that such a metric
is hyperk\"ahler and has a tri-holomorphic $U(1)$ isometry generated
by $\frac{\partial}{\partial t}.$ Since the harmonic function $\tau_2$
does not depend on $\chi,$ there is an additional $U(1)$ isometry generated
by the vector field $\frac{\partial}{\partial \chi}.$ It is easy to check
that it is also tri-holomorphic. Thus the asymptotic metric on the moduli
space has a tri-holomorphic $T^2$ isometry, as promised. Moreover, it looks
like a $T^2$-fibration over the $z$-plane, and the $T^2$ action is
fiberwise. Moreover, there is a distinguished complex structure on this
$T^2$ fibration, defined up to a sign, with respect to which the projection
map is holomorphic. This is the complex structure
\begin{eqnarray}\label{disting}
\frac{\partial}{\partial z}&\mapsto &-i\frac{\partial}{\partial z},\\ \nn
\frac{\partial}{\partial\bz}&\mapsto & i\frac{\partial}{\partial\bz},\\ \nn
\frac{\partial}{\partial t}&\mapsto & \frac{1}{\tau_2}\left(
\frac{\partial}{\partial\chi}-\tau_1(z)\frac{\partial}{\partial t}\right),\\
\frac{1}{\tau_2}\left(\frac{\partial}{\partial\chi}-
\tau_1(z)\frac{\partial}{\partial t}\right) &\mapsto &
-\frac{\partial}{\partial t}.\nn
\end{eqnarray}
The nice thing about the distinguished complex structure is that it can
be computed not only for well-separated monopoles, but everywhere on the
moduli space~\cite{usone}. The geometry of the resulting elliptic fibration
is discussed in detail in the next section.

The expressions Eqs.~(\ref{elliptic},\ref{tautwomon}) do not completely specify the asymptotic metric on the moduli space because we have not fixed the period of $t.$
Let the period be $2\pi/p,$ where $p\in (0,+\infty).$ To determine $p,$
we note that $\tau(z)$ is a multi-valued function of $z.$ This is not
so surprising, if we realize that, in the distinguished complex structure,
$p\tau(z)$ is the Teichm\"uller parameter of the $T^2$ fiber at point $z,$
which is only defined up to a $PSL_2(\ZZ)$ transformation. It is not hard to
verify that the metric is well-defined if and only if the monodromy of
$p\tau(z)$ belongs to $PSL_2(\ZZ).$
Here it is important to remember that $z,\chi,t$ are the relative coordinate
of two monopoles, and thus the points $(z,\chi,t)$ and $(-z,-\chi,-t)$
must be identified. Therefore $\tau(z)$ and $\tau(-z)$ must
be related by a $PSL_2(\ZZ)$ transformation. From Eq.~(\ref{tautwomon})
it follows that under $z\ra -z$ the monodromy is $\tau\ra\tau +1.$
This implies that $p\in\NN.$

The precise value for $p$ depends on the choice of the topology of the
gauge group. One can equally well work with an $SU(2)$ or an $SO(3)=SU(2)/\ZZ_2$ gauge group. This ambiguity has the following consequence. The coordinate $t$ on the moduli space parametrizes ``large'' gauge transformations which leave invariant the Higgs field at infinity. Such transformations form
a $U(1)$ subgroup of the gauge group. When one passes from an $SU(2)$
gauge group to its $\ZZ_2$ quotient, the period of $t$ reduces by a factor
$2,$ and therefore the value of $p$ increases by a factor $2.$ We will
see in the next section that when the gauge group is taken to be $SO(3),$
one has $p=4.$ Therefore if the gauge group is taken to be $SU(2)$ (the
more standard choice for non-abelian monopoles), we have $p=2,$ and
$t$ has period $\pi.$

The metric~(\ref{elliptic},\ref{tautwomon}) is applicable for large
$|z|.$ If we try to continue it formally to small $|z|,$ we encounter
a singularity at the hypersurface $\tau_2(z)=0.$ This singularity is
at a finite distance, so the metric ~(\ref{elliptic},\ref{tautwomon})
is geodesically incomplete. This is completely analogous to the
case of ordinary monopoles on $\RR^3:$ the exact metric on the relative
moduli space of two monopoles (the Atiyah-Hitchin metric) asymptotically
looks like a Taub-NUT metric with a ``wrong'' sign of the Taub-NUT parameter,
so that the naive continuation of the asymptotic metric is geodesically
incomplete. We expect that the exact metric on the relative
moduli space of two periodic monopoles is smooth and complete, just like
the Atiyah-Hitchin metric. However, the exact metric cannot have a
tri-holomorphic $U(1)$ isometry. This can be seen, for example, from the
fact that in the limit $\vr\ra\infty,$ when periodic monopoles
reduce to ordinary monopoles~\cite{ustwo}, the exact metric must reduce
to the Atiyah-Hitchin metric, which does not have continuous
tri-holomorphic isometries.

\subsection{Multi-Monopole Interactions}
It is obvious how to extend this procedure to interactions of several dyons.  The Lagrangian turns out to be
\begin{eqnarray}
\frac{1}{4\pi}L&=&\sum_{j=1}^k\left(-\frac{v}{2g}q_j^2+\frac{gv}{2}\vec{V}_j^2+b q_jV_{j\chi}\right)+\sum_{1\leq i< j\leq k}
\left(\frac{g^2}{2}u(z_i-z_j)(\V_i-\V_j)^2+\right.\nn\\
&&+\left.g w(z_i-z_j)(q_i-q_j)(V_{i\chi}-V_{j\chi})-\frac{1}{2} u(z_i-z_j)(q_i-q_j)^2\right).\nn
\end{eqnarray}
Using an identity
\begin{equation}
k\sum_{j=1}^k a_j b_j=\left(\sum_{j=1}^k a_j\right)\left(\sum_{j=1}^k b_j\right)+\sum_{1\leq i< j\leq k} (a_i-a_j)(b_i-b_j),
\end{equation}
we can rewrite $L$ as a sum of the center-of-mass  Lagrangian
\begin{equation}
\frac{1}{4\pi}L_{CM}= \frac{vg}{2k}\left(\sum_{j=1}^k
\vec{V}_j\right)^2-\frac{v}{2gk}\left(\sum_{j=1}^k
q_j\right)^2+\frac{b}{k}\left(\sum_{j=1}^k q_j\right)\left(\sum_{l=1}^k V_{l\chi}\right),
\end{equation}
and the Lagrangian describing the relative motion
\begin{eqnarray}\label{all}
\frac{1}{4\pi}L_{{\rm rel}}&=&\sum_{1\leq i< j\leq k}
\left\{\left(\frac{g\vr}{2k}+\frac{g^2}{2\pi}\log|z_i-z_j|\right)
(\V_i-\V_j)^2+\right.\nn\\
&&+\left(\frac{b}{k}+\frac{g}{\pi}\arg(z_i-z_j)\right)
(q_i-q_j)(V_{i\chi}-V_{j\chi})-\\
&&-\left.\left(\frac{\vr}{2gk}+\frac{1}{2\pi}\log|z_i-z_j|\right)
(q_i-q_j)^2\right\}.\nn
\end{eqnarray}
In the limit $N\ra\infty,v\ra\infty$ with $\vr$ and $b$ fixed, the
relative Lagrangian stays finite, while the center-of-mass Lagrangian
diverges.

The relative moduli space of $k$ monopoles on $\RR^2\times\SS^1$ has
the geometry of a $2(k-1)$-dimensional torus fibered over $\RR^{2k-2}$.
The torus is parameterized by monopoles' relative positions along $\SS^1$ and their relative phases, while the coordinates on the base are given by the monopoles' relative positions on $\RR^2$. The general form of the metric is
given by the expression
\begin{eqnarray}\label{fibre}
&&ds^2=\frac{1}{2}g_{ij}(dz_id\bar{z}_j+d\bar{z}_idz_j)+
\tilde{g}_{ij}d\chi_i d\chi_j+\\
&&h_{ij}\left[dt_i+W_{ik} d\chi_k+{\rm Re}\left(Z_{ik}d\bar{z}_k\right)\right]
\left[dt_j+W_{jl}d\chi_l+{\rm Re}\left(Z_{jl}d\bar{z}_l\right)\right],\nn
\end{eqnarray}
restricted to the submanifold $\sum_j z_j=\mu,$ $\sum_j\chi_j=\alpha$ and $\sum_j t_j=\beta$ for some constants $\mu,$
$\alpha,$ and $\beta.$ These restrictions are imposed to fix the position of the center of mass and the total phase. The variables $t_j,\chi_j$
are periodic with $j$-independent periods.

To determine the metric coefficients, we note that the asymptotic
metric must have $U(1)^{k-1}$ isometry acting fiberwise.
Without loss of generality, we may assume that the corresponding Killing
vector fields are given by
$$
\frac{\partial}{\partial t_{j+1}}-\frac{\partial}{\partial t_j},
\quad j=1,\ldots,k-1.
$$
Then all metric coefficients must be independent of $t_j.$ The corresponding
integrals of motion must be identified with $q_j.$ Thus we may
compute the reduced Lagrangian which is independent of $\dot{t}_j$
by performing the Legendre transform on $\dot{t}_j.$
We then compare with Eq.~(\ref{fibre}) and obtain the following
answer:
\begin{eqnarray}
\frac{1}{4\pi} g_{ii}&=&\frac{1}{4\pi}\tilde{g}_{ii}=g\vr\frac{k-1}{k}+\frac{g^2}{\pi}
\sum_{j\neq i}\log|z_i-z_j|,\nn\\
\frac{1}{4\pi}g_{ij}&=&\frac{1}{4\pi}\tilde{g}_{ij}=-\frac{g\vr}{k}-
\frac{g^2}{\pi}\log|z_i-z_j|,\ (i\neq j)\nn\\
W_{ii}&=&b\frac{k-1}{gk}+\frac{1}{\pi}
\sum_{j=i+1}^k\arg(z_i-z_j)+\frac{1}{\pi}\sum_{j=1}^{i-1}\arg(z_j-z_i),\nn\\
W_{ij}&=&-\frac{b}{gk}-\frac{1}{\pi}\arg(z_i-z_j),\ (i<j)\nn\\
W_{ij}&=&-\frac{b}{gk}-\frac{1}{\pi}\arg(z_j-z_i),\ (i>j)\nn\\
Z_{ij}&=&0,\nn\\
4\pi g^2 \left(h^{-1}\right)_{ii}&=&
\frac{\vr}{g}\frac{k-1}{k}+\frac{1}{\pi}
\sum_{j\neq i}\log|z_i-z_j|,\nn\\
4\pi g^2 \left(h^{-1}\right)_{ij}&=&
-\frac{\vr}{gk}-\frac{1}{\pi}\log|z_i-z_j|,\ (i\neq j).\nn
\end{eqnarray}
Note that the matrix $h^{-1}$ is not invertible. However, it is
invertible on the subspace defined by $\sum_j dt_j=\sum_j d\chi_j=0,$
and that is all we need. Similarly, the matrix $g_{ij}$ has a
one-dimensional kernel, but on the submanifold of interest it is
positive-definite if all $|z_i-z_j|$ are large.

This metric is very similar to the one found by Gibbons and
Manton for monopoles on $\RR^3$. They are both special cases of a common
ansatz (Eqs.~(23),(28),(29) of Ref.~\cite{GM}) which is the most general
$4(k-1)$-dimensional hyperk\"ahler metric with a tri-holomorphic $U(1)^{k-1}$
isometry~\cite{HKLR,PP}.
In our case all the metric coefficients are independent of $t_j,\chi_j,$ and therefore we have $2(k-1)$  commuting Killing vector fields
\begin{equation}\label{killing}
\frac{\partial}{\partial t_{j+1}}-\frac{\partial}{\partial t_j},\quad
\frac{\partial}{\partial\chi_{j+1}}-\frac{\partial}{\partial\chi_j},
\quad j=1,\ldots,k-1.
\end{equation}
It is easy to check that they are tri-holomorphic. Thus the asymptotic
metric on the relative moduli space admits a tri-holomorphic
$T^{2(k-1)}$ isometry.

It remains to fix the periodicity of the variables $t_{j+1}-t_j.$
For the metric given by Eq.~(\ref{fibre}) to be well-defined,
the period must be $2\pi/p,$ with $p\in\NN.$
When two of the monopoles are far from the rest, the
metric must agree with that found in the previous subsection.
This implies that $p$ is equal to $2$ or $4$ depending on whether the
gauge group is $SU(2)$ or $SO(3).$

The multi-monopole metric is valid when the separations $|z_j-z_j|$
between all the monopoles are large. If we try to continue the metric to
small separations, $g_{ij}$ ceases to be invertible on the submanifold of interest. The resulting singularities indicate that the asymptotic
metric is not geodesically complete. The exact metric is expected to be
smooth and complete.

\subsection{Two Monopoles with Singularities}

It is straightforward to derive the moduli space metric in the
presence of Dirac-type singularities on $\RR^2\times\SS^1$.
We will write it down only for the case $k=2$ (two periodic monopoles).
This case is of particular interest because the relative moduli space is
four-dimensional, and therefore one obtains new examples of gravitational
instantons. As explained in Ref.~\cite{ustwo}, the number of Dirac singularities
$n$ cannot exceed $2k=4,$ therefore we obtain five gravitational
instantons corresponding to $n=0,1,\ldots,4.$ We denote them $D_n$,
$n=0,\ldots,4.$ The reason for this nomenclature is the following.
Gravitational instanton of type $D_n$ is isomorphic to the Coulomb branch
of $N=2$ supersymmetric $SU(2)$ gauge theory with $n$ hypermultiplets
on $\RR^3\times\SS^1$~\cite{ustwo}. The latter theory has $SO(2n)$ global flavor symmetry in the ultraviolet.

At long distances the fields created by $n$ singularities and the two
nonabelian monopoles are in a $u(1)$ Cartan subalgebra of the $su(2)$ gauge algebra. (This $u(1)$ subalgebra is defined locally by the condition that it
leaves invariant the Higgs field.)
Each of the singularities has magnetic charge
$g_j=-1/2$~\cite{ustwo}, while each 't~Hooft-Polyakov
monopole has magnetic charge $1$~\cite{ustwo}.
Since the singularities are stationary and
have no electric charge, their only effect is to replace the constant
background fields $\vr$ and $b$ with $\vr+\sum_{j=1}^n g_j u(z_{1,2}-m_j)$ and
$b+\sum_{j=1}^n g_j w(z_{1,2}-m_j),$ respectively. Here $m_j,$ $j=1,\ldots,n,$
are the $z$-coordinates of the singularities and $z_{1,2}$ are respective positions of
the 't~Hooft-Polyakov monopoles.

The Lagrangian is given by
\begin{eqnarray}
\frac{1}{4\pi} L&=& \left(\frac{v}{2}-\frac{1}{4}\sum_{j=1}^n u(z_1-m_j)\right) \vec{V}_1^2+
\left(\frac{v}{2}-\frac{1}{4}\sum_{j=1}^n u(z_2-m_j)\right)\vec{V}_2^2\nn\\
&+&\frac{1}{2}u(z_1-z_2)\left(\vec{V}_1-\vec{V}_2\right)^2\nn\\
&-&\left(\frac{v}{2}-\frac{1}{4}\sum_{j=1}^n u(z_1-m_j)\right) q_1^2-
\left(\frac{v}{2}-\frac{1}{4}\sum_{j=1}^n u(z_2-m_j)\right)q_2^2\nn\\
&-&\frac{1}{2}u(z_1-z_2)\left(q_1-q_2\right)^2\nn\\
&+&\left(b-\frac{1}{2}\sum_{j=1}^n w(z_1-m_j)\right)q_1V_{1\chi}+
\left(b-\frac{1}{2}\sum_{j=1}^n w(z_2-m_j)\right)q_2V_{2\chi}\nn\\
&+&w(z_1-z_2)(q_1-q_2)\left(V_{1\chi}-V_{2\chi}\right).\nn
\end{eqnarray}
To get the Lagrangian describing the relative motion, we set $\vec{V}_1+\vec{V}_2=0$ and $q_1+q_2=0,$ and obtain:
\begin{eqnarray}
\frac{1}{4\pi}L_{rel}\!\!\!&=&\!\!\!\left(\frac{v}{4}+\frac{1}{2}u(z_1-z_2)-
\frac{1}{16}\sum_{j=1}^n(u(z_1-m_j)+u(z_2-m_j))\right)
\left(\vec{V}_1-\vec{V}_2\right)^2\nn\\
&+&\!\!\!\left(\frac{b}{2}+w(z_1-z_2)-\frac{1}{8}
\sum_{j=1}^n(u(z_1-m_j)+u(z_2-m_j))\right)\nn\\
&\times &\!\!\! (q_1-q_2)\left(V_{1\chi}-V_{2\chi}\right)\nn\\
&-&\!\!\!\left(\frac{v}{4}+\frac{1}{2}u(z_1-z_2)-\frac{1}{16}
\sum_{j=1}^n(u(z_1-m_j)+u(z_2-m_j))\right)(q_1-q_2)^2.\nn
\end{eqnarray}

>From this expression we immediately see that the divergent constant $C_N$
in the function $u(z)$ can be absorbed into a renormalization of $v$:
$$
v_{ren}=v-\frac{4-n}{2} C_N.
$$
This is precisely the same renormalization which makes the Higgs field
$\phi$ finite in the limit $N\ra\infty$ with fixed $v_{ren}.$

We can also read off the metric on the relative moduli space.
As in subsection(\ref{2mon}), we introduce the relative coordinates
$z=z_1-z_2, \chi=\chi_1-\chi_2, t=t_1-t_2,$ and set $z_1+z_2=0.$ (More generally, we could set $z_1+z_2=c$ for some $c\in\CC,$ but the constant
$c$ can always be absorbed into a shift of $m_i,$ so one does not gain
anything by considering non-zero
$c$.) The resulting asymptotic metric has the form Eq.~(\ref{elliptic})
with the function $\tau(z)$ given by
\begin{equation}\label{tausing}
\tau(z)=i\left(\frac{\vr-ib}{2}+\frac{1}{\pi}\log(\bar{z})-
\frac{1}{8\pi}\sum_{j=1}^n\log(\bar{m}_j^2-\frac{1}{4}\bar{z}^2)\right).
\end{equation}
In particular, for $n=4$ $\tau(z)$ has a trivial monodromy around
$z=\infty.$

This metric is valid when non-abelian monopoles are far from each other
and the Dirac monopoles, i.e. when $|z|$ and $|z\pm 2m_i|,$ $i=1,\ldots,n,$
are all large.

Unlike in the previous cases, there is no ambiguity in the choice of
gauge group here. Recall that the magnetic
charge of the Dirac singularity is $-1/2.$ The geometric meaning of
this non-integral magnetic charge is that the monopole bundle on
$\RR^2\times\SS^1$ is an $SO(3)$ bundle which cannot be lifted
to an $SU(2)$ bundle~\cite{ustwo}. The obstruction is the second
Stiefel-Whitney class evaluated on a sphere centered at the Dirac
singularity. Thus only $SO(3)$ is a consistent choice of the gauge group.

The monodromy of $\tau$ around
the points $z=\pm 2 m_i$ is given by $\tau\ra\tau + 1/4.$ On the other
hand, if the period of $t$ is $2\pi/p,$ then the monodromy of
$p\tau$ must be in $PSL_2(\ZZ).$ This implies that $p/4\in\NN.$
The minimal value for $p$ is $4,$ in which case $t$ has
period $\pi/2.$
In the next section we will show that in the presence of Dirac
singularities the minimal choice $p=4$ is the right one.
This is also the right value for $p$ in the absence of singularities,
provided that the gauge group is taken to be $SO(3).$

\section{Geometry of New Gravitational Instantons}

In the previous section we have constructed five gravitational instantons
($D_n,$ $n=0,\ldots,4$)
and showed that they are ALG manifolds. In this section we discuss their
topology and geometry.

The basic observation is that the distinguished complex structure
on the moduli spaces of periodic monopoles is easy to compute using the
monopole spectral curve defined in Refs.~\cite{usone,ustwo}.
Let us specialize the
results of Refs.~\cite{usone,ustwo} to the present case. To each
solution of the $U(2)$ Bogomolny equations (possibly with singularities)
one can associate an algebraic curve in $\CC\times \CC^*$. If the number
of non-abelian monopoles is $2$, and the $z$-coordinates of the singularities
are given by $m_i,i=1,\ldots,4,$ then the curve has the form
$$
(y-m_1)(y-m_2)w^2+a(y^2-u)w+b(y-m_3)(y-m_4)=0.
$$
Here $y\in\CC$, $w\in\CC^*$ are the coordinates in $\CC\times\CC^*$, and the parameters $a,b\in\CC^*$ can be expressed in terms of the asymptotic behavior of the monopole fields, see Ref.~\cite{ustwo}.
The complex parameter $u$ is the modulus of the curve (i.e. it is not fixed
by the boundary conditions on the monopole fields).
Thus there is a map from the monopole moduli space $X$ to the complex
$u$-plane. As explained in the above-cited papers, this map is holomorphic (in the distinguished complex structure), and its
fiber is the Jacobian of the curve. Since the curve is elliptic in our case,
the fiber coincides with the curve itself. It follows that $X$ is an
elliptic fibration over $\CC$. The asymptotic
coordinate $z$ of the previous section should be identified with $\sqrt u$
times a constant factor. We will see below that with our normalizations
this constant factor is unity.

It is helpful to note that this elliptic fibration is precisely
the Seiberg-Witten fibration for the $N=2,$ $d=4$ gauge theory with gauge group $SU(2)$ and four fundamental hypermultiplets with masses
$m_i,i=1,\ldots,4.$\footnote{As explained in Refs.~\cite{usone,ustwo}, this coincidence follows from very general string-theoretic considerations. In fact
Seiberg-Witten solutions for many gauge theories can be derived by considering
periodic monopoles for various gauge groups.}
This is trivial to see if we use the form of the Seiberg-Witten fibration
found in Ref.~\cite{Witten}. Thus we can borrow the results
in the physics literature~\cite{SW1,SW2,APSW} on the geometry of this fibration.

For generic $m_i$ there are six singular fibers each of which is a rational curve with a node (i.e. a singular curve $y^2=x^2$). In Kodaira's classification of singularities of elliptic fibrations~\cite{Ko},
these are type-$I_1$ singular fibers. The Euler characteristic of an
$I_1$ fiber is $1$, so the Euler characteristic of $X$ is $6$. It is easy to see that $b_1(X)=b_3(X)=b_4(X)=0,$ hence $b_2(X)=5$.

When all $m_i$ are large, four out of six singularities occur
near $u=m_i^2,$ i.e. far out in the moduli space. In this region of
the moduli space the asymptotic formula~(\ref{disting}) for the distnguished
complex structure is valid and should agree with the results obtained
from the spectral curve approach. Indeed, we see from Eq.~(\ref{tausing})
that $\tau(z)$ has four singularities at $z=m_i.$ Thus in the
asymptotic region $u\simeq z^2,$ as claimed. Moreover, this comparison
allows us to infer the precise
periodicity of the coordinate $t$ left undetermined by the analysis of the
previous section. There, we saw that if the period of $t$ is $2\pi/p,$
then for large $|z|$ the Teichm\"uller parameter of the elliptic fiber at point $z$ is $p\tau(z),$ where $\tau(z)$ is given by
Eq.~(\ref{tausing}). The asymptotic metric is well-defined if
$p/4\in\NN.$ From Eq.~(\ref{tausing}) we see that the monodromy of
$p\tau(z)$ near $z=m_i$ is such that the singularity is
of type $I_{p/4}.$
Therefore agreement with the spectral curve approach requires $p=4.$

If one sets all $m_i$ to zero,
then all six singular fibers coalesce into a single singular fiber
at $u=0$, and the $j$-invariant of the curve becomes $u$-independent. The singularity at $u=0$ is of type
$I_0^*$ in Kodaira's classification. This means
that the singular fiber is a union of five rational curves whose intersection
matrix is the affine Cartan matrix of type $D_4.$ Since $b_2(X)=5,$
these rational curves span $H_2(X)$, and therefore the
intersection form on $H_2(X)$ is the affine $D_4$ Cartan
matrix.\footnote{Since $X$ is non-compact, the intersection form need
not be non-degenerate. In the present case, the kernel of the intersection form is one-dimensional.} From the viewpoint of the quantum $SU(2)$ gauge theory, the $I_0^*$
singularity corresponds to a non-trivial CFT with global
$SO(8)$ symmetry~\cite{SW2}.

In general, the elliptic fibration corresponding to the $D_4$ ALG manifold
can have 1,3,4,5, or 6 singular fibers~\cite{SW2,APSW}. The types of singular
fibers that can occur are given by the following list:
$$
I_0^*, I_1,\ I_2,\ I_3,\ I_4,\ II,\ III,\ IV.
$$
We have already discussed the physical meaning of $I_0^*$
singularity. $I_n$ singularity corresponds to the infrared behavior
of $N=2$ $U(1)$ gauge theory with $n$ massless charge-1
hypermultiplets~\cite{SW2}. Singular fibers of type $II,$ $III,$ and $IV$ correspond to non-trivial CFTs (so-called Argyres-Douglas points)~\cite{APSW}.
It is more convenient to use the notation $H_1,H_2,$ and
$H_3,$ respectively for these singularities.

Now let us discuss the geometry of the remaining ALG manifolds ($D_n$
with $0\leq n\leq 3$).
It is easy to see what happens when we decrease the number of Dirac
monopoles $n$ by taking one or more $m_i$ to infinity. $X$ is still
elliptically fibered over $\CC$, but the number of singular fibers is now
given by $n+2$ for generic $m_i$. Each of the singular fibers is of type $I_1$. It follows that the Euler characteristic is $n+2$, and the second Betti number is $n+1$. By Zariski's lemma, the self-intersection number
of each singular fiber vanishes, and therefore the rank of the intersection form is at most $n$. In particular, for $n=0$ the second
homology is one-dimensional and the intersection form vanishes
altogether.

For $n<4$ it is impossible to tune the remaining $m_i$ to bring all $n+2$
singular fibers together~\cite{APSW}. At most one can bring $n+1$ of them together, so that the elliptic fibration has two singular fibers.
It has been shown in Ref.~\cite{APSW} that one of them is an
$I_1$ singularity, while the other one is of type $I_1$, $H_1 (II)$,
$H_2 (III)$, or $H_3 (IV)$,
depending on whether $n=0,1,2,$ or $3$. More generally, an elliptic
fibration corresponding to the ALG manifold of type $D_n$ may have from
2 to $n+2$ singular fibers. The types of singular fibers that occur are
$I_\ell$ and $H_\ell$, $1\leq \ell\leq n$.

Note that the intersection form of a singular fiber of type $I_n$
has rank $n-1$ (it is the affine $A_{n-1}$ Cartan matrix). Hence
we may conclude that the rank of the intersection form on $H_2(X)$
is either $n-1$ or $n.$ We saw above that for $n=0$ the intersection form
vanishes identically, while for $n=4$ it coincides with the affine Cartan
matrix of type $D_4.$ It would be interesting to compute the
intersection form for the remaining cases ($n=1,2,3$).

\section{Realization via Hitchin equations}

In Refs.~\cite{usone,ustwo} we showed that Nahm transform establishes a one-to-one
correspondence between periodic monopoles with Dirac singularities and
solutions of Hitchin equations on a cylinder with particular boundary
conditions. Furthermore, we showed that the map between the corresponding
moduli spaces is bi-holomorphic if one uses the natural complex structures.
Both moduli spaces are hyperk\"ahler manifolds, and analogy with the case
of monopoles on $\RR^3$ suggests that Nahm transform induces an isometry
between them. If this is true, then we have an alternative
construction of ALG gravitational instantons using the moduli space of
Hitchin equations.

Let us focus on the case of ALG manifold of type $D_4$, when
the boundary conditions for Hitchin equations are especially simple.
According to Ref.~\cite{ustwo}, if the number of Dirac singularities is
four and the number of non-abelian monopoles is two, then the Hitchin data
consist of a $U(2)$ connection $\hA$ and a Higgs field $\hphi$ on a cylinder
$\RR\times\SS^1$ with two points removed. We will identify $\RR\times\SS^1$
with a strip $0\leq {\rm Im} s < 1$ in a complex $s$-plane (this requires picking an orientation on $\RR\times\SS^1$). The two punctures
are located at $s=s_1$ and $s=s_2$.
Away from the punctures $\hA$ and $\hphi$ satisfy the Hitchin
equations~\cite{HitchinSpec}
$$
\bpartial_\hA\hphi=0,\quad
\hF_{s\bs}+\frac{i}{4}\left[\hphi,\hphi^\dag\right]=0,
$$
while near the punctures they have the following behavior:
$$
\hphi(s)\sim \frac{R_i}{s-s_i},\quad \hA_s\sim \frac{Q_i}{s-s_i},\quad i=1,2.
$$
Here $R_i$ and $Q_i$ are rank-one matrices which can be simultaneously
diagonalized by a gauge transformation. Their eigenvalues depend on the
behavior of the monopole fields near $z=\infty,$ see Ref.~\cite{ustwo} for details.
We should also specify the behavior of $\hphi$ and $\hA$ at infinity.
Let $r={\rm Re}\ s$. Let $m_i\in\CC, i=1,\ldots,4,$ be the $z$-coordinates
of the Dirac singularities, and $\chi_i\in \RR/(2\pi\ZZ)$ be their
coordinates on $\SS^1$.
For $|r|\ra\infty$ the connection $A$ becomes flat; the asymptotic holonomy is given by
$$
\diag\left(e^{i\chi_1},e^{i\chi_2}\right)
$$
for $r\ra -\infty$ and
$$
\diag\left(e^{i\chi_3},e^{i\chi_4}\right)
$$
for $r\ra +\infty$.
The eigenvalues of the Higgs field tend to $m_1,m_2$ for $r\ra-\infty$
and to $m_3,m_4$ for $r\ra +\infty$.

It is rather obvious that the moduli space of Hitchin equations is a hyperk\"ahler manifold. Indeed, Hitchin equations on a
cylinder can be regarded as moment map equations for an infinite-dimensional
HKQ. The quotienting group is the group of gauge transformations, and
it acts on the cotangent bundle of an infinite-dimensional affine
hyperk\"ahler space, the space of $U(2)$ connections on a cylinder.
The residues of the Higgs field and the connection at $s_1,s_2$ can
be regarded as the level of the moment map.

Solutions of Hitchin equations of this kind have been extensively studied by C.~Simpson~\cite{Simpson} and others. To make explicit the connection
with Simpson's work, we make a conformal transformation $w=e^{2\pi s}$
which maps the cylinder with two punctures into a $\PP^1$ with four
punctures, with $w$ being the coordinate on the North patch of the $\PP^1$.
Hitchin equations are conformally invariant if we agree that
$\hphi$ transforms as a 1-form, i.e. $\hphi_s ds=\hphi_w dw$.
Then $\hphi_w$ has simple poles at all four punctures, with residues which
can be simultaneously diagonalized by a gauge transformation. Furthermore,
$\hA_w$ also has simple poles, and the residues can be diagonalized
simultaneously with the residues of $\hphi_w$.

One can simplify this further
by noting that the trace and traceless parts of $\hA_w,\hphi_w$ separately
satisfy Hitchin equations. Hitchin equations for the trace part simply
say that $\Tr\ \hphi_w$ and $\Tr\ \hA_w$ are holomorphic 1-form and
flat connection on a punctured $\PP^1$, and therefore are completely
determined by their residues. Thus the moduli space
of $U(2)$ Hitchin equations can be replaced by the moduli space of
$SU(2)$ Hitchin equations. Using the results of Ref.~\cite{ustwo}, one can easily
compute the eigenvalues of the residues of the $SU(2)$ Higgs field and connection in terms of locations of the Dirac singularities and the asymptotic behavior of the monopole fields. In the notation of Ref.~\cite{ustwo}, the eigenvalues of the residues of the Higgs field at the four punctures are given by
$$
\pm\frac{1}{2}(m_1-m_2),\ \pm\frac{1}{2}(m_3-m_4),\ \pm\frac{\mu_1}{2},\
\pm\frac{\mu_2}{2},
$$
and the eigenvalues of the residues of the connection are given by
$$
\pm\frac{i}{2}(\chi_1-\chi_2),\ \pm\frac{i}{2}(\chi_3-\chi_4),\
\pm\frac{i\alpha_1}{8\pi},\ \pm\frac{i\alpha_2}{8\pi}.
$$
The parameters $\mu_1, \mu_2$ and $\alpha_1, \alpha_2$ can be expressed
in terms of the asymptotic behavior of the monopole fields~\cite{ustwo}.

When the number of Dirac singularities is less than four, the Nahm transform
is again given in terms of solutions of Hitchin equations on $\PP^1.$
But the singularities of $\hphi$ and $\hA$ are more complex in this case
(they are not ``tame,'' in the terminology of Ref.~\cite{Simpson}). For more
details, see Ref.~\cite{ustwo}.

\section*{Acknowledgments}
S.~Ch. was supported in part by NSF grant PHY9819686. A.~K. was supported
in part by DOE grants DE-FG02-90-ER40542 and DE-FG03-92-ER40701.

\end{document}